# ON THE MOTION OF FREE MATERIAL TEST PARTICLES IN ARBITRARY SPATIAL FLOWS[*]

Tom Martin
Gravity Research Institute
Boulder, Colorado 80306-1258
tmartin@rmi.net

## Abstract

*We show how the motion of free material test particles in arbitrary spatial flows is easily determined within the context of ordinary vector calculus. This may be useful for everyone, including engineers and other non-specialists, when thinking about gravitational problems. It already has valid application to simple problems such as the problems of motion in rotating and accelerating frames and to the gravitational problem of the single spherically symmetric attractor. When applied to the two body gravitational problem, it may help us determine the actual direction of the flow.*

## Introduction

In a recent publication [1], we discussed the possibility that Nature might prefer non-static and spatially flowing type solutions of the gravitational field equations rather than the usual static and spatially curved type solutions. In the case of the two body gravitational problem of the Earth-Sun system, we discovered that there is a very strong gravitational time dilation effect near the gravitational saddle point which can be utilized, experimentally, to distinguish the physical reality of the two possible types of solution.

It is our intention in the present paper to show how the equations of motion of free material test particles in arbitrary spatial flows can be easily determined within the context of ordinary vector calculus. This obviates the need to introduce the usual complication of covariant differentiation with its associated Christoffel symbols and affine parameters. The resulting General Relativistic equations of motion already have valid application to simple problems such as the problems of motion in rotating and accelerating frames of reference and to the gravitational problem of the single spherically symmetric attractor. When applied to the two body gravitational problem, they may help us determine the actual direction of the flow.

---

[*] http://xxx.lanl.gov/ftp/gr-qc/papers/9807/9807006.pdf

## 1. Space-time with Spatial Flow

When space-time is characterized by a flow of physical space, there exists a global Galilean coordinate frame $\{\mathbf{r},t\}$ in which the flow is represented by a 3-space vector field $\mathbf{w} \equiv \mathbf{w}(\mathbf{r},t)$. This flow of space is a generalization of Newton's concept of absolute space in which no part of space is moving with respect to any other part (Newton's absolute space corresponds to the case in which there is a Galilean frame in which $\mathbf{w} \equiv \mathbf{0}$ everywhere). In these Galilean coordinates, the proper time element of an atomic clock is given by

$$d\boldsymbol{t} = \boldsymbol{g}^{-1} dt \equiv \sqrt{1 - u^2/c^2}\, dt \quad , \tag{1-1}$$

where

$$\mathbf{u} \equiv \mathbf{v} - \mathbf{w} \quad . \tag{1-2}$$

Here, $\boldsymbol{t}$ is the proper time of the clock, $c$ is the coordinate speed of light in physical space (a constant), $t$ is the coordinate time, $\mathbf{r}$ is the coordinate position vector of the clock, $\mathbf{v} \equiv d\mathbf{r}/dt$ is the coordinate velocity of the clock, and $\mathbf{u}$ is the coordinate velocity of the clock relative to physical space.

From the above equations, we see that the space-time line element is

$$c^2 d\boldsymbol{t}^2 = (c^2 - w^2)dt^2 + 2\mathbf{w}\cdot d\mathbf{r}\, dt - (d\mathbf{r})^2 \tag{1-3}$$

in these Galilean coordinates. Furthermore, the total time dilation of a clock in motion with velocity $\mathbf{v}$ (we restrict $u < c$) in an arbitrary spatial flow $\mathbf{w}$ is given by

$$d\boldsymbol{t} = \sqrt{1 - (v^2 + w^2 - 2\mathbf{v}\cdot\mathbf{w})/c^2}\, dt \quad . \tag{1-4}$$

We have shown in reference [1] that these Galilean coordinates are perfectly acceptable coordinates for studying 1) the General Relativistic problem of the rotating frame in flat space-time and 2) the General Relativistic problem of the single spherically symmetric attractor. Thus, the equations of motion obtained in the following two Sections are applicable to these *canonical* cases as well as to many others.

## 2. The Equations of Motion

The path of a *free* material test particle in General Relativity is one over which the integral of the particle's proper time is an extremum:

$$d \int dt = 0 \ . \qquad (2\text{-}1)$$

The path is obtained from the associated Euler-Lagrange equations. In Galilean coordinates with spatial flow $\mathbf{w}$, the calculation of the path is simplified by using the coordinate time $t$ as the path parameter. This reduces the number of Euler-Lagrange equations from four (involving the space-time coordinates) to three (involving the spatial coordinates). From (1-1), we have

$$d \int d\boldsymbol{t} = d \int c^{-1}(c^2 - u^2)^{1/2} \, dt = 0 \ . \qquad (2\text{-}2)$$

Let us perform the calculation in rectangular Galilean coordinates $\{x, y, z, t\}$ with the orthonormal spatial basis $\{\hat{\mathbf{e}}_x, \hat{\mathbf{e}}_y, \hat{\mathbf{e}}_z\}$. The Euler-Lagrange equations are

$$\frac{d}{dt}\left(\frac{\partial L}{\partial \mathbf{v}}\right) = \frac{\partial L}{\partial \mathbf{r}} \ , \qquad (2\text{-}3)$$

where

$$L = L(\mathbf{r}, \mathbf{v}; t) = c^{-1}(c^2 - \mathbf{u} \cdot \mathbf{u})^{1/2} = \mathbf{g}^{-1} \qquad (2\text{-}4)$$

is the Lagrangian, and

$$\frac{\partial L}{\partial \mathbf{v}} \equiv \frac{\partial L}{\partial v^x}\hat{\mathbf{e}}_x + \frac{\partial L}{\partial v^y}\hat{\mathbf{e}}_y + \frac{\partial L}{\partial v^z}\hat{\mathbf{e}}_z \ , \qquad (2\text{-}5)$$

$$\frac{\partial L}{\partial \mathbf{r}} \equiv \partial_x L \hat{\mathbf{e}}_x + \partial_y L \hat{\mathbf{e}}_y + \partial_z L \hat{\mathbf{e}}_z \equiv \mathbf{grad}\, L \quad . \tag{2-6}$$

In this Euler-Lagrange formalism, $\mathbf{r}$ and $\mathbf{v}$ are treated as independent variables. $\mathbf{v} \equiv \mathbf{v}(t)$ is a parametric vector, while $\mathbf{w} \equiv \mathbf{w}(\mathbf{r},t)$ is a vector field. Thus,

$$\mathbf{grad}\, \mathbf{u} = \mathbf{grad}\,(\mathbf{v} - \mathbf{w}) = -\mathbf{grad}\, \mathbf{w} \quad , \tag{2-7}$$

where

$$\mathbf{grad}\, \mathbf{w} \equiv \begin{matrix} +\partial_x w^x \hat{\mathbf{e}}_x \hat{\mathbf{e}}_x & +\partial_x w^y \hat{\mathbf{e}}_x \hat{\mathbf{e}}_y & +\partial_x w^z \hat{\mathbf{e}}_x \hat{\mathbf{e}}_z \\ +\partial_y w^x \hat{\mathbf{e}}_y \hat{\mathbf{e}}_x & +\partial_y w^y \hat{\mathbf{e}}_y \hat{\mathbf{e}}_y & +\partial_y w^z \hat{\mathbf{e}}_y \hat{\mathbf{e}}_z \\ +\partial_z w^x \hat{\mathbf{e}}_z \hat{\mathbf{e}}_x & +\partial_z w^y \hat{\mathbf{e}}_z \hat{\mathbf{e}}_y & +\partial_z w^z \hat{\mathbf{e}}_z \hat{\mathbf{e}}_z \end{matrix} \tag{2-8}$$

is the tensor characterizing the spatial inhomogeneity of the flow $\mathbf{w}$.

From (2-7), we have

$$-\mathbf{grad}(\mathbf{u} \cdot \mathbf{u}) = -2(\mathbf{grad}\, \mathbf{u}) \cdot \mathbf{u} = 2(\mathbf{grad}\, \mathbf{w}) \cdot \mathbf{u} \quad ,$$

hence

$$\frac{\partial L}{\partial \mathbf{r}} = \mathbf{grad}\, c^{-1}(c^2 - \mathbf{u} \cdot \mathbf{u})^{1/2} = \frac{1}{2} L^{-1} c^{-2} \mathbf{grad}\,(c^2 - \mathbf{u} \cdot \mathbf{u}) = L^{-1} c^{-2} (\mathbf{grad}\, \mathbf{w}) \cdot \mathbf{u} \quad .$$

Since

$$L = c^{-1}(c^2 - \mathbf{u} \cdot \mathbf{u})^{1/2} = c^{-1}(c^2 - \mathbf{v} \cdot \mathbf{v} + 2\mathbf{w} \cdot \mathbf{v} - \mathbf{w} \cdot \mathbf{w})^{1/2} \quad ,$$

$$\frac{\partial L}{\partial \mathbf{v}} = \frac{1}{2} L^{-1} c^{-2} (2\mathbf{w} - 2\mathbf{v}) = -L^{-1} c^{-2} \mathbf{u} \quad ,$$

so

$$\frac{d}{dt}\left(\frac{\partial L}{\partial \mathbf{v}}\right) = -c^{-2}\left(\frac{dL^{-1}}{dt}\right)\mathbf{u} - c^{-2}L^{-1}\frac{d\mathbf{u}}{dt} = c^{-4}L^{-3}\left(\mathbf{u}\cdot\frac{d\mathbf{u}}{dt}\right)\mathbf{u} - c^{-2}L^{-1}\frac{d\mathbf{u}}{dt} \ .$$

Setting $L^{-2} = \mathbf{g}^2$ (from (2-4)), we see that our Euler-Lagrange equations (2-3) are

$$\frac{d\mathbf{u}}{dt} + \frac{\mathbf{g}^2}{c^2}\left(\mathbf{u}\cdot\frac{d\mathbf{u}}{dt}\right)\mathbf{u} = -(\mathbf{grad}\,w)\cdot\mathbf{u} \ . \tag{2-9}$$

Taking the scalar product of this equation with $\mathbf{u}$, we get

$$\frac{d\mathbf{u}}{dt}\cdot\mathbf{u} + \frac{\mathbf{g}^2}{c^2}\left(\mathbf{u}\cdot\frac{d\mathbf{u}}{dt}\right)(\mathbf{u}\cdot\mathbf{u}) = -((\mathbf{grad}\,w)\cdot\mathbf{u})\cdot\mathbf{u} \ . \tag{2-10}$$

Since $\mathbf{u}\cdot\mathbf{u} = c^2(1-\mathbf{g}^{-2})$, (2-10) becomes

$$\mathbf{g}^2\,\frac{d\mathbf{u}}{dt}\cdot\mathbf{u} = -((\mathbf{grad}\,w)\cdot\mathbf{u})\cdot\mathbf{u} \ . \tag{2-11}$$

Substituting this back into (2-9), we find that

$$\frac{d\mathbf{u}}{dt} = -(\mathbf{grad}\,w)\cdot\mathbf{u} + \frac{1}{c^2}(((\mathbf{grad}\,w)\cdot\mathbf{u})\cdot\mathbf{u})\,\mathbf{u} \ . \tag{2-12a}$$

We can write this more formally as

$$\boxed{\frac{d\mathbf{u}}{dt} + (\mathbf{1} - (1/c^2)\mathbf{u}\mathbf{u})\cdot(\mathbf{grad}\,w)\cdot\mathbf{u} = 0} \ , \tag{2-12b}$$

where $\mathbf{1}$ is the identity tensor in 3-space and $\mathbf{1} - (1/c^2)\mathbf{u}\mathbf{u}$ is the symmetric *tensor of motion* of the test particle.

This is the equation of motion of a *free* material test particle in Galilean coordinates with arbitrary spatial flow $\mathbf{w}$. The equation is *fully relativistic* (in the sense that the material test particle may have an arbitrary relativistic speed ($0 \le u < c$) relative to physical space).

# 3. Beyond Newton

The *non-relativistic* (slow motion) approximation of the relativistic equation (2-12) is

$$\frac{d\mathbf{u}}{dt} = -(\mathbf{grad\,w})\cdot\mathbf{u} \quad. \tag{3-1a}$$

In terms of $\mathbf{v}$ and $\mathbf{w}$, this is

$$\frac{d\mathbf{v}}{dt} - \frac{d\mathbf{w}}{dt} = -(\mathbf{grad\,w})\cdot(\mathbf{v}-\mathbf{w}) \quad, \tag{3-1b}$$

where $d\mathbf{w}/dt$ is the derivative of the parametric vector $\mathbf{w}(t) = \mathbf{w}(\mathbf{r}(t),t)$ along the path $\mathbf{r}(t)$ specified by the velocity vector $\mathbf{v} = d\mathbf{r}/dt$. By the chain rule of differentiation,

$$\frac{d\mathbf{w}}{dt} = \mathbf{v}\cdot\mathbf{grad\,w} + \partial_t\mathbf{w} \quad. \tag{3-2}$$

Using the vector identities

$$\mathbf{v}\cdot\mathbf{grad\,w} - (\mathbf{grad\,w})\cdot\mathbf{v} = (\mathbf{curl\,w})\times\mathbf{v} \tag{3-3}$$

and

$$(\mathbf{grad\,w})\cdot\mathbf{w} = \frac{1}{2}\mathbf{grad}\,w^2 \quad, \tag{3-4}$$

we find that

$$\frac{d\mathbf{v}}{dt} = \frac{1}{2}\mathbf{grad}\,w^2 + (\mathbf{curl\,w})\times\mathbf{v} + \partial_t\mathbf{w} \quad. \tag{3-5}$$

To remind ourselves that the acceleration $\mathbf{a} \equiv d\mathbf{v}/dt$ is that of a *free* material test particle, we write this as

$$\boxed{\mathbf{a}_{free} = \frac{1}{2}\mathbf{grad}\,w^2 + (\mathbf{curl\,w})\times\mathbf{v} + \partial_t\mathbf{w}} \quad. \tag{3-6}$$

The first term on the right-hand side of this equation represents the gravitational and generalized centrifugal accelerations, the second term represents the generalized Coriolis acceleration, and the third term is the acceleration arising from an explicit dependence of the flow on time.

This equation provides us with an easy way to study the problem of motion in rotating and accelerated frames in ordinary Newtonian mechanics. We do this by focusing on the flows *induced* by the frame motions. For example, a rotating Galilean frame in flat space-time with an axial rotational velocity $\boldsymbol{\omega} \equiv \boldsymbol{\omega}(t)$, induces the flow

$$\mathbf{w}(\mathbf{r},t) \;=\; \mathbf{r} \times \boldsymbol{\omega}(t) \;. \tag{3-7}$$

More importantly, equation (3-6) has application *beyond* ordinary Newtonian mechanics, because it holds for *arbitrary* flows and not just those induced by the motion of frames. An example of a flow that is not induced by the motion of a frame is the flow associated with a single spherically symmetric gravitational attractor. In a spherical Galilean frame centered on the attractor, this is given by one or the other of the flows

$$\mathbf{w}(\mathbf{r}) = \pm\sqrt{2GM/r}\,\hat{\mathbf{e}}_r \;, \tag{3-8}$$

where $G$ is the gravitational constant and $M$ is the mass of the attractor. As we discussed in some detail in reference [1], the Principle of General Covariance assures us that no physical experiment can distinguish between the two flows (3-8) as long as we are dealing with a single *isolated* attractor. However, when a second attractor is present, there are terms in (3-6) which may help us in designing an experiment to measure the actual direction of the flow. We discuss this possibility, at least qualitatively, in the next Section.

# 4. Satellite Motion through the Surface of Transition of the Earth-Sun System

Suppose a satellite with a stable atomic clock onboard is sent through the region of the Earth-Sun gravitational saddle point as suggested in reference [1]. If the frequency of the clock varies in a way that is consistent with the spatial flow type solution of the two body problem, we will be justified in studying the details of this flow in other ways.

Let us discuss the Earth-Sun two body flow from the point of view of a non-rotating Galilean frame centered on the Sun. In this frame of reference, the Earth and the gravitational saddle point are orbiting about the Sun. (We neglect all other planetary perturbations.) The flow is quiescent at the saddle point [1], and there will be a surface of transition between the Earth's spatial flow and the Sun's spatial flow. In the region of the saddle point, this hyperboloid-like surface is swept away from the Sun, and it is moving with the Earth and the saddle point as a boundary between the terrestrial flow and the solar flow. It is closest to the Earth where it intersects the saddle point (the saddle point is about 260,000 km from the Earth).

Consider a point $P$ on the surface of transition and in the orbital plane at a distance of several thousand kilometers from the saddle point. At some, but as yet experimentally undetermined, small distance into the terrestrial side of the flow from $P$, the terrestrial flow will be approximately parallel to the transition surface and the orbital plane, and it will have a speed on the order of $w = \sqrt{2Gm/r} \cong 1.75$ km/sec, where $m$ is the Earth's mass and $r$ is the distance from the Earth to the saddle point. On the solar side of the flow from $P$, the solar flow will also be approximately parallel to the transition surface and the orbital plane, but its speed will be on the order of $w = \sqrt{2GM/R} \cong 42.2$ km/sec, where $M$ is the solar mass and $R$ is the distance from the Sun to the saddle point.

If we send a satellite from the Earth perpendicularly through the surface of transition at $P$ and out into the solar flow, we may be able to study its motion with equation (3-6). This depends on the nature of the transition between the flows. There are three possibilities: 1) the transition is a strict tangential discontinuity of the flow, 2) the transition is turbulent flow, or 3) the transition is a smoothly varying flow. We can apply equation (3-6) with confidence only to the third case.

Suppose, for the sake of argument, that the flow is smoothly varying through the transition. In view of the differences of the flows on either side of $P$ as estimated above, the term $(1/2)\mathbf{grad}\, w^2$ would give the satellite an acceleration in its forward direction, while the terms $(\mathbf{curl}\,\mathbf{w}) \times \mathbf{v}$ and $\partial_t \mathbf{w}$ would give the satellite accelerations *in the* *direction of the solar flow*. So, in the case of a smoothly varying transition, it is possible, in principle, to determine the actual direction of the flow.